\documentclass[letterpaper,journal]{IEEEtran}
\usepackage{amsmath,amsfonts,amssymb}
\usepackage{algorithm}
\usepackage{algpseudocode}
\usepackage{array}
\usepackage[caption=false,font=normalsize,labelfont=sf,textfont=sf]{subfig}
\usepackage{textcomp}
\usepackage{stfloats}
\usepackage{url}
\usepackage{verbatim}
\usepackage{graphicx}
\usepackage{cite}
\usepackage{booktabs}
\usepackage[table]{xcolor}
\usepackage{colortbl}
\usepackage{multirow}
\usepackage{capt-of}
\usepackage{makecell}
\usepackage[T1]{fontenc}
\usepackage[hidelinks]{hyperref}
\graphicspath{{Fig/}}

\definecolor{myblue}{RGB}{66,133,244}
\definecolor{myred}{RGB}{219,68,55}
\definecolor{mypurple}{RGB}{156,39,176}
\definecolor{bestgreen}{RGB}{220,245,220}
\definecolor{failred}{RGB}{250,225,225}

\begin{document}

\title{Topology-only Local Network Alignment, Part 2. DijSLA: Stochastic Greedy Extension for Cross-Species GO Annotation Prediction}

\author{Yiran~Jin,
        Tingyin~Ding$^{\dagger}$,
        Utsav~Jain$^{\dagger}$,
        Grady~Jaron~Yu$^{\ddagger}$,
        Burwin~Liu$^{\ddagger}$,
        and~Wayne~B.~Hayes$^{\ast}$%
\thanks{Y. Jin is with the Department of Electrical Engineering and Computer Science, University of California, Irvine, Irvine, CA 92697 USA (e-mail: yiranj12@uci.edu).}%
\thanks{T. Ding, U. Jain, G. J. Yu, and B. Liu were with the Department of Computer Science, University of California, Irvine, Irvine, CA 92697 USA.}%
\thanks{W. B. Hayes is with the Department of Computer Science, University of California, Irvine, Irvine, CA 92697 USA (e-mail: whayes@uci.edu).}%
\thanks{$^{\dagger}$These authors contributed equally. $^{\ddagger}$These authors contributed equally.}%
\thanks{$^{\ast}$Corresponding author: Wayne B. Hayes.}}
\markboth{IEEE Transactions on Computational Biology and Bioinformatics}%
{Jin \MakeLowercase{\textit{et al.}}: Topology-only Local Network Alignment, Part 2}

\maketitle

\begin{abstract}
Sequence similarity has long been a strong signal for protein function prediction, whereas the predictive value of network topology has been difficult to verify. Since a protein's function is largely defined by its interaction partners, and since similar interactions are expected across species, aligning protein--protein interaction (PPI) networks by topology alone should uncover functionally related proteins. Topology-only {\it global} network alignment has recently produced high-quality functional predictions, but it has yet to be shown that local regions of topological similarity can do the same. We introduce DijSLA (Dijkstra-inspired Stochastic Local Alignment), the first {\it local} network alignment algorithm designed from the ground up to use only topology. We adopt a seed-and-extend strategy: starting from seed pairs of proteins with similar network topology, we expand the alignment in a Dijkstra-inspired manner, following edges out of the aligned set to node pairs with near-maximal similarity. Rather than expanding deterministically, we inject randomness so that we can sample many high-quality alignments, in which node pairs with robust topological similarity ``rise above'' the noise and reveal functional similarity. Using networks and GO annotations from April 2010, we predict annotations validated against those from April 2020. With Network Alignment Frequency (NAF), a robustness measure that correlates with prediction precision, we make predictions with precision on par with or better than the best sequence-based methods and broadly similar to that of global topological alignment. Compared with existing local aligners, DijSLA produces larger, topologically high-quality alignments, which yield dramatically more predictions with higher validation rates, while using less CPU time and far less memory. Code is available at \url{https://github.com/waynebhayes/BLANT/tree/master/CDijkstra}.
\end{abstract}


\section{Introduction and Motivation}

\IEEEPARstart{P}{roteins} play a central role in all aspects of cellular processes, including metabolism, molecular transport, and structural organization \cite{braun2012history}. Consequently, understanding protein function is fundamental to modern biomedicine, pharmacology, and systems biology \cite{ofran2005beyond,chautard2009interaction,rao2014protein,lin2024comprehensive}.  However, experimental characterization of protein function remains labor-intensive and costly. As a result, computational annotation of protein function has emerged as a crucial research area to address the growing gap between available biological protein data and experimentally validated functional knowledge \cite{radivojac2013large,dhanuka2023comprehensive, lin2024comprehensive}.

A protein’s amino acid sequence dictates the three-dimensional structure into which it folds \cite{whisstock2003prediction,petsko2004protein}, along with the cellular environment in which the fold occurs \cite{hartl2011molecular}. This structure then governs how the protein interacts with other proteins and how the complexes it participates in interact with other complexes, which determines its function \cite{fersht1999structure}.

Similarity at each stage of this process correlates with similarity in function and can be used for annotation. Sequence-based annotation remains a major approach, as whole-genome sequencing and high-throughput experiments have greatly expanded available sequence data. Pairwise sequence alignment tools such as FASTA \cite{pearson1988improved}, BLAST \cite{altschul1990basic} and PSI-BLAST \cite{altschul1997gapped} infer function from E-value \cite{zehetner2003ontoblast, hawkins2009pfp} or from sequence identity and coverage \cite{piovesan2011bar}; profile-based methods use motifs, domains \cite{rentzsch2013protein} and functional residues \cite{wass2008confunc, gong2016gofdr}, at the higher cost of building multiple sequence alignments; and machine learning methods infer function from sequence similarity in a learned feature space \cite{lobley2008ffpred, saracc2010gopred, cao2021tale}.

However, the path from sequence to function is long and error-prone: implicitly or explicitly, we need to step through the processes from sequence to transcription to folding to binding domains to interactions to  functions; an error at any step can cause an error in functional prediction. Protein–protein interaction (PPI) networks, instead, provide a more direct approach to function prediction by giving the complexes and interaction partners of each protein. This has led to a vast body of work on topology-based PPI network analysis. %
Single-network analysis is based on the idea that interacting proteins are more likely to share the same function than not\cite{schwikowski2000network}. Typical methods use Markov chains \cite{hishigaki2001assessment,deng2002prediction} or cluster and community detection \cite{nepusz2012detecting,manipur2021community}. 

Beyond single-network analysis, comparative network analysis is an important perspective, since many proteins across species may share similar functions through evolution, which can be reflected as conserved patterns in their PPI networks \cite{matthews2001identification,sharan2005conserved,memivsevic2010complementarity}. To identify such conserved functional modules, network alignment approaches hold great promise \cite{wang2022current}.

Network alignment algorithms usually require a guide signal. Very few network alignment methods are capable of producing alignments of high enough quality to make functional predictions using topology alone; SANA \cite{wang2022sana} is one that uses global alignment; most or all other methods also rely on sequence similarity \cite{ciriello2012alignnemo,mina2012alignmcl, neyshabur2013netal, micale2014gasoline,kalecky2018primalign, maskey2019leprimalign}. This motivates a central question: can PPI network topology alone provide meaningful signal for cross-species protein function prediction?

Indeed, sequence information is a powerful signal for protein function annotation. Yet topology provides complementary but vital information as has been demonstrated in \cite{bandyopadhyay2006systematic,kuchaiev2010topological,memivsevic2010complementarity}. Ultimately, the ideal framework for protein function annotation would integrate all available biological information, including sequence, structure, and PPI network topology. However, each individual source of information should first pass a \textit{litmus test} of prediction capability on its own before being considered a reliable factor \cite{wang2022current}. For topology, this requires removing sequence completely, so that its strong and well-understood signal does not overwhelm the weak yet vital topological one. Very few global methods are designed to use network topology alone and to the best of our knowledge, DijSLA is the first \textit{local} network alignment method to show that network topology alone can predict protein function annotations at scale with high validation rates.

Machine learning methods have also been explored in recent years, using neural networks to embed PPI networks and assess protein similarity in the embedded space for function predictions. For example, DeepGO \cite{kulmanov2018deepgo} and MultipredGO \cite{giri2020multipredgo} employ convolutional neural networks (CNNs), DeepGOA \cite{zhang2020deep} utilizes long short-term memory (LSTM) networks and KOGAL \cite{djeddi2025degree} uses knowledge graph embeddings. However, all of these combine network embeddings with sequence-based features and are designed to use both. Removing sequence falls outside their intended use: without retraining on topology alone and even altering their architectures to remove sequence, their performance would not reflect the value of topology, and such retraining is beyond the scope of this paper. In network alignment, by contrast, sequence can generally be added or removed in a controlled way without altering the algorithm. Further, the interpretability of machine learning models is inherently weak, whereas in network alignment each decision is supported by explicit topology evidence, which makes it possible to study directly how topology relates to function.

\subsection{Local Network Alignment vs. Global Network Alignment}\label{subsec1}

Similar to sequence alignment, network alignment can be local or global. Local network alignment aims to align small, highly similar regions that may be conserved across evolution. Merge-and-mine \cite{guzzi2018survey} style methods first construct a large alignment graph of protein pairs guided by sequence similarity or orthology, which usually requires a huge amount of RAM, then search the networks for well-aligned neighboring pairs to build the alignment. Examples include NetworkBLAST \cite{sharan2005conserved,kalaev2008networkblast}, Mawish \cite{koyuturk2006pairwise}, AlignNemo \cite{ciriello2012alignnemo} and AlignMCL \cite{mina2012alignmcl}. Network topology here plays only a secondary role, and these algorithms may fail to produce any alignment if deprived of sequence information. LePrimAlign \cite{maskey2019leprimalign} also falls into this category, using PageRank scores from PrimAlign \cite{kalecky2018primalign}, which merges the networks first via sequence similarity.

Mine-and-merge \cite{guzzi2018survey} methods reverse that: they first identify regions in each network separately and then align them together. Such methods include: MOPHY \cite{erten2009phylogenetic}, Jancura's method \cite{jancura2012methodology}, and MODULA \cite{guzzi2015modula}, which detect clustered complexes first within each network and then align these clusters together by assessing their similarity. Our method is closer to mine-and-merge, since we mine topology-based node-pair similarities, then merge via seed-and-extend. Note that the above methods are all pairwise network alignment. There are also methods such as GASOLINE \cite{micale2014gasoline} capable of multi-network alignment.

Global alignment, in contrast, aims to align entire networks to maximize overall similarity, giving a more complete view of the evolutionary relationship between them. The common strategy is to optimize a cost function of the entire alignment. Methods include IsoRank \cite{singh2008global,liao2009isorankn}, which maximizes a convex combination of the sequence similarity and the size of the common subgraph by framing it as an eigenvalue problem, L-GRAAL \cite{malod2015graal}, which uses Lagrangian optimization for its starting point and SANA \cite{mamano2017sana}, which uses simulated annealing to optimize any given objective.

Meng et al.~\cite{meng2016local} proposed a fair comparison method of local network alignment and global network alignment. Although it was a good comparison at the time, both global and local network alignment algorithms have improved in the past decade. On the other hand, we agree with the survey that local alignment and global alignment discover different biological information. We contend here that—just as in the case of sequence—local and global network alignment are fundamentally different problems and are not directly comparable. In both cases, a global alignment effectively builds a rigid, inflexible ``structure'' that could mask local regions of similarity that occur in conflicting regions globally. This is why most sequence comparison today has converged on using local methods like BLAST rather than global sequence comparisons. A prime example in the sequence context is duplicate regions, where a long sequence may appear once in species A but more than once in species B; any global alignment is forced to ``choose'' which of the duplicates in B will be mapped to the one occurrence in A. It is clear that an analogous problem can happen in the case of ``duplicate'' network regions. For these reasons, we believe that local alignments are best discovered independently of global ones.

\begin{figure}[t]
    \centering
     
    \includegraphics[alt = {Diagram showing three example graphs, G1 (blue) and G2 (red), and their aligned graph. Matched node pairs are highlighted, conserved edges are shown in purple, and unmatched edges remain blue or red, illustrating how EC and S3 scores are computed.}, width=0.55\linewidth]{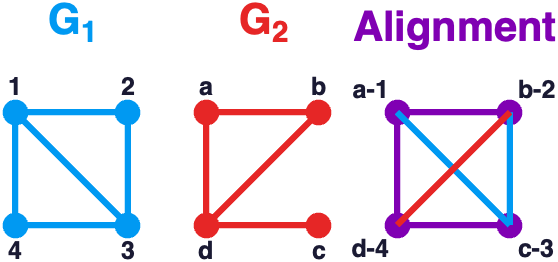}

    \caption[Network alignment and EC and S$^3$]{%
      Illustration of network alignment and computation of EC and $\mathbf{S^3}$ scores.
      Two example graphs, $G_1$ (\textcolor{myblue}{blue}) and $G_2$ (\textcolor{myred}{red}), with the pairs
      $a$--$1$, $b$--$2$, and $d$--$4$ already aligned. Here, \textcolor{mypurple}{purple} edges denote conserved interactions, and \textcolor{myblue}{blue} and \textcolor{myred}{red} edges are unconserved edges in $G_1$ and $G_2$, respectively. Counting the edges of each color in the rightmost (Alignment) network after aligning $c$--$3$, the edge conservation scores are
      $\mathrm{EC}_1=\tfrac{\text{\textcolor{mypurple}{purple}}}{\text{\textcolor{mypurple}{purple}}+\text{\textcolor{myblue}{blue}}} = \tfrac{3}{5}$ for $G_1$ and
      $\mathrm{EC}_2=\tfrac{\text{\textcolor{mypurple}{purple}}}{\text{\textcolor{mypurple}{purple}}+\text{\textcolor{myred}{red}}} = \tfrac{3}{4}$ for $G_2$,
      and the symmetric substructure score is
      $S^3=\tfrac{\text{\textcolor{mypurple}{purple}}}{\text{\textcolor{mypurple}{purple}}+\text{\textcolor{myblue}{blue}}+\text{\textcolor{myred}{red}}} = \tfrac{3}{6}$.
      }
    \label{fig:EC}
\end{figure}

\section{Methods}
\subsection{Seed-and-Extend}

Our method was initially inspired by the seed-and-extend approach of BLAST \cite{altschul1990basic}, adapted to the network context via an analogy to Dijkstra’s algorithm \cite{dljkstra1959note}. A seed is a pair of nodes—one from each network—that are aligned with high confidence using a local topology-based measure—sequence is not used even for seeds, although they {\it could} be. Starting from the seed, the alignment is “extended” by choosing a new pair of nodes that (i) have high topological similarity with each other, and (ii) are adjacent to the two nodes of an already-aligned pair, each in its own network. In contrast to Dijkstra’s algorithm, which grows deterministically, we intentionally inject noise so that the extension selects at random among top-scoring candidates. In addition, we impose an EC\footnote{Although everybody agrees on the mathematical definition of EC, and that E stands for Edge, there is no universal agreement on what the C stands for: Correctness, Conservation, Coverage, among others.} threshold (see Fig.~\ref{fig:EC}) $\epsilon$ at every extension step to ensure the growing alignment maintains high topological quality.

In this work, we use purely topological similarity measures, namely Importance \cite{hashemifar2014hubalign} and graphlet-based similarities, for both seeding and extension. Our alignments and predictions are therefore independent of sequence similarity, which enables biological predictions beyond the reach of sequence similarity. The Importance score measures how central a node is to the structure of its network. It is computed by repeatedly removing peripheral (degree 1) nodes and passing their weight to their neighbors, so that weight propagates inward from the periphery. The score therefore reflects a node’s global role in the network rather than only its local connectivity, and nodes with similar scores in two networks may play similar roles, suggesting they should be aligned. The graphlet-based score quantifies the local topological environment of a node by counting its participation in small, connected induced subgraphs known as graphlets \cite{prvzulj2004modeling}. The distance between graphlet degree vectors measures the similarity between the two nodes. We evaluated several variants and found that L-GRAAL’s measure \cite{kuchaiev2010topological,malod2015graal} works well. Other variants of graphlet-based similarity failed to produce meaningful alignments with any fixed top-similarity seed.

To seed, we adopted two approaches:
\begin{enumerate}
    \item Select the highest-scoring pair
    \item Select at random from near-top high-scoring pairs
\end{enumerate}
Approach 1 deterministically selects the highest-scoring pair from the global similarity list computed with the chosen topological similarity measure, ensuring that the alignment begins consistently with the best match. However, the most similar pair is not necessarily the best starting point, both because networks contain high amounts of noise \cite{han2005effect,wodak2013protein} which causes uncertainty in the true topology, and because the most similar pair does not guarantee potential for extension. This motivates Approach 2, in which we randomly select one seed pair from a set of near-top high-scoring pairs. This yields more varied alignments than always starting from the same top pair, while still ensuring that each alignment begins from a high-quality seed. In both approaches, the seed is a single pair of nodes, though multiple-pair seeds are also supported and can be explored, along with other seeding strategies, as introduced in our companion paper \cite{wang2026topology}.

Starting from the seed, the alignment is extended by exploring the direct neighbors of all nodes in the current alignment. For each aligned pair, we take the Cartesian product of its two nodes’ neighbor sets in the two networks; the union of these products over all aligned pairs forms the candidate set $\mathcal{C}$, from which we select one pair with high topological similarity to add to the alignment. To keep the alignment one-to-one, once a pair is added, all other candidates involving either of its nodes are removed from $\mathcal{C}$. As in seeding, the pair to add can be selected either deterministically, as the highest-scoring candidate, or with randomness, from a set of near-top high-scoring candidates. Furthermore, at each step, we check $\mathrm{EC}_1$ and $\mathrm{EC}_2$ against the preset EC threshold $\epsilon$ and discard the selected pair if either falls below $\epsilon$. 
Extension continues until no suitable candidate pairs remain.

\subsection{Stochasticity via the \texorpdfstring{$\delta$}{delta} Parameter}
Stochasticity is a key component of our algorithm, for seeding and especially for extension. To control it, we introduce a similarity threshold $\delta$, defined as the score tolerance below the highest-scoring pair. This threshold sets how far the selection may deviate from the greedy choice: a target score $s_t$ is drawn uniformly at random from $[s_{\max}-\delta,\ s_{\max}]$, and the candidate whose score is closest to $s_t$ is selected. This allows variation at each extension step. When $\delta = 0$, this reduces to greedy selection and the algorithm is deterministic. Running the algorithm multiple times with a small positive $\delta$ therefore yields many different high-quality alignments, each built from near-top-scoring pairs at every step. These alignments can differ substantially from each other, especially when they start from different seeds. 

Having many alignments allows us to assess the confidence of each aligned protein pair: a pair that recurs across many alignments is more likely to be ``correct''. Wang et al.~\cite{wang2022sana} formalized this idea as the Network Alignment Frequency (NAF), the number of alignments in which a given pair appears, and showed that NAF correlates with biological information. Following their approach, we run DijSLA $N$ times and compute the NAF of each aligned pair across the $N$ resulting one-to-one alignments; the maximum possible NAF is thus $N$. A high NAF is unlikely to arise by chance: under completely random selection, the probability that a given pair has NAF $= K$ is on the order of $\left(\frac{1}{|V_1|\cdot|V_2|}\right)^K$. Our alignments are subject to additional constraints, such as node connectivity, the preference for high-similarity pairs, and the EC threshold $\epsilon$, so this probability is higher in practice. Nevertheless, the principle remains: pairs that appear more often than expected by chance are more likely to reflect a strong signal, which helps counterbalance the noise and biases in PPI networks and allows these pairs to be used to predict functional relationships. As we show in Section~\ref{sec: GO Pred}, NAF indeed correlates with Gene Ontology (GO) annotation prediction: as NAF increases, prediction precision also tends to increase.

Selecting the candidate closest to a randomly drawn target score requires the candidates to be kept sorted by score, with fast lookup of any given score. The candidate set also changes at every step: new pairs are added from the neighbors of the newly aligned nodes, and pairs involving those nodes are removed. Because the candidate set grows quickly as the alignment expands, we store it in a skip list \cite{pugh1990skip}, which supports search, insertion and deletion all in expected $O(\log n)$ time.

\subsection{Detailed Algorithm Description}
\noindent Given graphs \(G_1=(V_1,E_1)\), \(G_2=(V_2,E_2)\), assume that $|V_1|\le |V_2|$.
A {\it local alignment} $\mathcal{A}$ is a one-to-one mapping from some subset nodes of $V_1$
to nodes in $V_2$; it is a set of ordered pairs $(v_1,v_2)$ with $v_1\in V_1, v_2\in V_2$ in
which no node from either network appears more than once.
For every possible pair $(v_1,v_2)$,
let \(M_{v_1v_2}\) be its similarity score under the chosen topological similarity measure (Importance or L-GRAAL). %

Algorithm~\ref{algo:local_alignment} describes the higher level of the DijSLA algorithm in pseudocode.
We start by choosing a {\it seed pair} of aligned nodes $(s_1,s_2)$, preferably with a relatively high similarity---within $\delta$ of the highest similarity node pair.
The local alignment is thus initialized as $\mathcal{A} = {\{(s_1,s_2)\}}$; the sets of nodes from $V_1$ and $V_2$ that have been aligned so far are, respectively, $U_1=\{s_1\}$ and $U_2=\{s_2\}$. We define the {\it candidate list} $\mathcal{C}$ as the set of all node pairs $(c_1,c_2)$ such that $c_i\in V_i\setminus U_i$ and, for some aligned pair $(u_1,u_2)\in\mathcal{A}$, $c_i$ is a neighbor of $u_i$ for both $i=1,2$---that is, the candidate list is the set of node-pairs that are ``one step'' away from the same aligned pair:
\[
\mathcal{C} = \bigcup_{(u_1,u_2)\in\mathcal{A}} \bigl(\mathcal{N}(u_1) \setminus U_1\bigr) \times \bigl(\mathcal{N}(u_2) \setminus U_2\bigr),
\]
where $\mathcal{N}(v)$ represents the set of neighbors of node $v$. Adding any candidate therefore creates at least one conserved edge, the one to the aligned pair it is adjacent to.

The alignment $\mathcal{A}$ is {\it extended} by sorting the set of candidate pairs $\mathcal{C}$ by similarity (highest-to-lowest), and then picking a high-scoring pair---specifically, some pair within $\delta$ of the highest-scoring pair in the candidate list, subject to the constraint that the new pair must not cause either $\mathrm{EC}_1$  or $\mathrm{EC}_2$ of the alignment (see\ Fig.~\ref{fig:EC}) to fall below the threshold $\epsilon$. Once the new pair $(c_1,c_2)$ is chosen and added to $\mathcal{A}$, all candidate pairs that include either $c_1$ or $c_2$ are removed from $\mathcal{C}$, and new candidate pairs adjacent to $(c_1,c_2)$ are added to $\mathcal{C}$. This process continues until $\mathcal{C}=\emptyset$, or until a user-defined number of consecutive candidate pairs have been discarded for falling below the EC threshold $\epsilon$, suggesting that no suitable candidates remain. The latter rule applies only after a minimum number of iterations (500 in our experiments) to ensure the alignment expands first; in practice, nearly all alignments terminated because $\mathcal{C}$ became empty, and EC-based discards were relatively infrequent.

\begin{algorithm}[!t]
\caption{DijSLA}
\label{algo:local_alignment}
\begin{algorithmic}[1]

\Require $G_1$, $G_2$; $M$; $(s_1,s_2)$; $\epsilon$; $\delta$

\State $\mathcal{A} \gets \{(s_1,s_2)\}$ \Comment{first aligned pair, the ``seed''}
\State $\texttt{alignSize} \gets 1$ \Comment{number of pairs in the alignment}
\State $U_1 \gets \{s_1\}$ \Comment{nodes used from $V_1$}
\State $U_2 \gets \{s_2\}$ \Comment{nodes used from $V_2$}
\State $\mathcal{C} \gets \big(\mathcal{N}(s_1)\setminus U_1\big)\times
                \big(\mathcal{N}(s_2)\setminus U_2\big)$ \Comment{candidate set}
\State $\texttt{skiplist} \gets \emptyset$
\State $\texttt{skiplist.insert}(\mathcal{C})$  \Comment{$\mathcal{C}$ sorted by similarity, highest at top}
\State $\texttt{discardCount} \gets 0$ \Comment{stop if too many discards}

\While{$\neg \texttt{skiplist.empty()}$ \textbf{and} $\texttt{discardCount} < \texttt{maxDiscard}$}
     \Comment{pop randomly a candidate pair within $\delta$ of top}
     \State $(c_1,c_2) \gets \texttt{skiplist.pop}(\delta)$
     \State $\mathcal{A} \gets \mathcal{A} \cup \{(c_1,c_2)\}$
    \If{$\mathrm{EC}_1(\mathcal{A}) \geq \epsilon$ \textbf{and} $\mathrm{EC}_2(\mathcal{A}) \geq \epsilon$}
        \State $U_1 \gets U_1 \cup \{c_1\}$;\;\; $U_2 \gets U_2 \cup \{c_2\}$
        \State $\texttt{++alignSize} $
        \State $\texttt{discardCount} \gets 0$
        \State $\texttt{skiplist.deleteAllContaining}((c_1,c_2))$
        \State $\Delta\mathcal{C} \gets \big(\mathcal{N}(c_1)\setminus U_1\big)\times
                        \big(\mathcal{N}(c_2)\setminus U_2\big)$ 
        \State $\texttt{skiplist.insert}(\Delta\mathcal{C})$ 
        \Comment{$\mathcal{C} \gets \mathcal{C} \cup \Delta\mathcal{C}$}
    \Else
        \State $\mathcal{A} \gets \mathcal{A} \setminus \{(c_1,c_2)\}$
        \State $\texttt{++discardCount}$
    \EndIf
\EndWhile

\end{algorithmic}
\end{algorithm}

\section{Results}
The ultimate test of a scientific method is its ability to predict {\em future} events, by which we mean events that have literally {\em never been seen} at the time the prediction is made. Thus, instead of using standard cross-validation, we apply a far more stringent test: we take networks available as of April 2010 (the first release of BioGRID V3 as shown in Table \ref{tab:biogrid_summary}), construct alignments across \textit{S. cerevisiae}, \textit{C. elegans}, \textit{D. melanogaster}, \textit{A. thaliana}, \textit{S. pombe}, \textit{M. musculus}, and \textit{R. norvegicus} with \textit{H. sapiens}. We take annotations from the Gene Ontology (GO) release of the same month and then use our alignments to predict annotations that were not seen in any form as of April 2010; we validate our predictions using annotations available 10 years later in April 2020.\footnote{We choose the exact same 10-year span as used in the global alignments in \cite{wang2022sana}, for easy comparison between local and global alignment results.} Note that by “unseen in any form”, we mean that we disregard functional predictions produced by our method if they were available {\em in any form}—e.g., using sequence-based prediction—in April 2010. This guarantees that our predictions are both genuinely novel, and not attainable using any other method available as of the prediction date.

\begin{table}[t]
\caption{Input PPI Networks.\protect\\ \upshape Networks from BioGRID-3.0.064 (April 2010) \cite{stark2010biogrid}, sorted by number of edges.}
\label{tab:biogrid_summary}
\centering
\resizebox{\columnwidth}{!}{%
\begin{tabular}{l c l r r r r}
\toprule
\textbf{Species} & \textbf{Abbreviation} & \textbf{Common name} &
\textbf{Nodes} & \textbf{Edges} & \textbf{Mean degree} & \textbf{Max degree} \\
\midrule
\textit{S.\ cerevisiae}    & SC & Baker’s yeast & 5611 & 51860 & 18.49 & 2552 \\
\textit{H.\ sapiens}      & HS & Human         & 8192 & 28117 & 6.86  & 284  \\
\textit{D.\ melanogaster} & DM & Fruit fly     & 7003 & 22128 & 6.32  & 176  \\
\textit{C.\ elegans}      & CE & Round worm    & 2794 & 4457  & 3.19  & 181  \\
\textit{A.\ thaliana}     & AT & Water cress   & 1607 & 2563  & 3.19  & 57   \\
\textit{S.\ pombe}        & SP & Fission yeast & 1374 & 2323  & 3.38  & 439  \\
\textit{M.\ musculus}     & MM & Mouse         & 528  & 475   & 1.80  & 25   \\
\textit{R.\ norvegicus}   & RN & Rat           & 232  & 178   & 1.53  & 8    \\
\bottomrule
\end{tabular}%
}
\end{table}

For each pair of species, we independently generate $N = 100$ local alignments using DijSLA to compute NAF for prediction. For each similarity measure, and for each change in a global parameter such as $\delta$ or $\epsilon$, we generate a new set of 100 alignments. To compare DijSLA with existing local aligners---AlignMCL, AlignNemo, LePrimAlign and GASOLINE---we likewise run each of them 100 times. All methods are restricted to topological information: we replace their sequence-based or orthology-based similarity with either Importance or L-GRAAL similarity, while keeping all other parameters at their default values. Other methods like MOPHY, Jancura's method and MODULA do not yet have publicly available code. Machine learning methods are not included, as removing sequence information requires altering their model architectures and retraining, which is beyond the scope of this paper.

Each set of 100 runs is executed as a Slurm job array of 100 tasks, with each task allocated a single CPU, on a cluster of nodes with 24-core Intel(R) Xeon(R) CPU X5680 @3.33GHz. We then compare the methods on these runs in terms of (i) computational efficiency, measured by runtime and peak memory usage; (ii) alignment quality, measured by alignment size, EC and $S^3$; and (iii) functional predictive performance, measured by the number and precision of the GO annotation predictions. DijSLA outperforms all other methods in virtually every important metric: it creates bigger topologically high-quality local alignments; these in turn provide dramatically more predictions with significantly higher validation rates, all while consuming less CPU time and far less RAM.

\begin{table}[t]
\centering
\caption{Size of the Input Similarity Files.\protect\\ \upshape Number of node pairs in the similarity files for the Importance (Imp) and L-GRAAL similarity measures. --: not available, owing to RN’s sparsity.}
\label{tab:simfilesize}
\footnotesize
\setlength{\tabcolsep}{5pt}
\renewcommand{\arraystretch}{1.15}

\begin{tabular}{lrr}
\toprule
\textbf{Species} &
\multicolumn{2}{c}{\textbf{Number of node pairs}} \\
\cmidrule(lr){2-3}
& \textbf{Imp} & \textbf{L-GRAAL} \\
\midrule
SC--HS & 3,558,408 & 45,965,312 \\
DM--HS & 1,629,860 & 57,368,576 \\
CE--HS & 203,912   & 22,888,448 \\
AT--HS & 114,880   & 13,164,544 \\
SP--HS & 93,340    & 11,255,808 \\
MM--HS & 2,872     & 4,325,376 \\
RN--HS & --        & 1,900,544 \\
\bottomrule
\end{tabular}
\end{table}

\begin{table*}[t]
\centering
\caption{Runtime and Peak Memory Usage of the Local Alignment Methods.\protect\\ \upshape Averages over 100 runs for each species pair with Importance or L-GRAAL similarity. Green highlights the best performance, and red indicates out-of-memory failures, for which runtime and memory are measured until the failure. AlignMCL's L-GRAAL runs on SC, AT, SP, MM and RN failed on a single memory allocation; all other failures were killed by the out-of-memory killer.}
\label{tab:runtime_memory_all}

\resizebox{\textwidth}{!}{
\begin{tabular}{ll|ccccc|ccccc}
\hline

\multirow{2}{*}{\textbf{Species}} &
\multirow{2}{*}{\textbf{Sim.}} &
\multicolumn{5}{c|}{\textbf{Runtime}} &
\multicolumn{5}{c}{\textbf{Peak Memory (GiB)}} \\

&
& \makecell{Align\\MCL} & \makecell{Align\\Nemo} & \makecell{LePrim\\Align} & GASOLINE & \makecell{DijSLA\\(ours)}
& \makecell{Align\\MCL} & \makecell{Align\\Nemo} & \makecell{LePrim\\Align} & GASOLINE & \makecell{DijSLA\\(ours)}\\

\hline

\multirow{2}{*}{SC--HS}
& Imp
& \cellcolor{failred}11.4\,min & \cellcolor{failred}2.13\,d & 1.45\,h & 1.17\,min & \cellcolor{bestgreen}\textbf{41.0\,s}
& \cellcolor{failred}90.3 & \cellcolor{failred}49.6 & 3.8 & 2.9 & \cellcolor{bestgreen}\textbf{0.21} \\

& L-GRAAL
& \cellcolor{failred}35.9\,min & \cellcolor{failred}2.77\,h & 50.0\,min & 4.88\,min & \cellcolor{bestgreen}\textbf{4.63\,min}
& \cellcolor{failred}35.7 & \cellcolor{failred}72.2 & 48.6 & 26.9 & \cellcolor{bestgreen}\textbf{0.21} \\

\hline

\multirow{2}{*}{DM--HS}
& Imp
& \cellcolor{failred}17.2\,min & \cellcolor{failred}2.39\,d & 1.18\,min & 37.5\,s & \cellcolor{bestgreen}\textbf{15.4\,s}
& \cellcolor{failred}77.4 & \cellcolor{failred}49.3 & 1.64 & 2.4 & \cellcolor{bestgreen}\textbf{0.25} \\

& L-GRAAL
& \cellcolor{failred}2.17\,h & \cellcolor{failred}3.65\,h & 56.3\,min & 6.75\,min & \cellcolor{bestgreen}\textbf{3.24\,min}
& \cellcolor{failred}78.9 & \cellcolor{failred}73.0 & 60.0 & 32.6 & \cellcolor{bestgreen}\textbf{0.25} \\

\hline

\multirow{2}{*}{CE--HS}
& Imp
& \cellcolor{failred}38.5\,min & \cellcolor{failred}3\,d & 12.8\,s & 7\,s & \cellcolor{bestgreen}\textbf{3.25\,s}
& \cellcolor{failred}78.1 & \cellcolor{failred}66.2 & 0.29 & 1.3 & \cellcolor{bestgreen}\textbf{0.10} \\

& L-GRAAL
& \cellcolor{failred}1.13\,h & \cellcolor{failred}2.97\,h & 25.8\,min & 2.14\,min & \cellcolor{bestgreen}\textbf{41.1\,s}
& \cellcolor{failred}90.9 & \cellcolor{failred}64.5 & 24.3 & 15.5 & \cellcolor{bestgreen}\textbf{0.11} \\

\hline

\multirow{2}{*}{AT--HS}
& Imp
& \cellcolor{failred}3.37\,h & \cellcolor{failred}2.37\,d & 6.7\,s & 4.13\,s & \cellcolor{bestgreen}\textbf{2.52\,s}
& \cellcolor{failred}74.1 & \cellcolor{failred}71.5 & 0.19 & 1.03 & \cellcolor{bestgreen}\textbf{0.064} \\

& L-GRAAL
& \cellcolor{failred}17.1\,min & \cellcolor{failred}2.92\,h & 12.9\,min & 1.03\,min & \cellcolor{bestgreen}\textbf{23.7\,s}
& \cellcolor{failred}24.6 & \cellcolor{failred}59.0 & 13.9 & 8.9 & \cellcolor{bestgreen}\textbf{0.064} \\

\hline

\multirow{2}{*}{SP--HS}
& Imp
& \cellcolor{failred}6.5\,h & \cellcolor{failred}3.5\,d & 1.58\,min & 4.79\,s & \cellcolor{bestgreen}\textbf{1.93\,s}
& \cellcolor{failred}49.2 & \cellcolor{failred}72.7 & 0.17 & 0.85 & \cellcolor{bestgreen}\textbf{0.056} \\

& L-GRAAL
& \cellcolor{failred}9.73\,min & \cellcolor{failred}7.67\,h & 11.2\,min & 1.13\,min & \cellcolor{bestgreen}\textbf{18.0\,s}
& \cellcolor{failred}17.0 & \cellcolor{failred}59.7 & 12.0 & 8.1 & \cellcolor{bestgreen}\textbf{0.058} \\

\hline

\multirow{2}{*}{MM--HS}
& Imp
& 10.9\,s & 59.1\,s & 1.82\,min & \cellcolor{bestgreen}\textbf{1.1\,s} & 1.57\,s
& 0.089 & 2.17 & 0.077 & 0.16 & \cellcolor{bestgreen}\textbf{0.029} \\

& L-GRAAL
& \cellcolor{failred}3.13\,min & \cellcolor{failred}10.2\,h & 5.99\,min & 21.4\,s & \cellcolor{bestgreen}\textbf{6.55\,s}
& \cellcolor{failred}5.71 & \cellcolor{failred}57.5 & 4.58 & 3.0 & \cellcolor{bestgreen}\textbf{0.029} \\

\hline

RN--HS
& L-GRAAL
& \cellcolor{failred}1.17\,min & \cellcolor{failred}7.68\,h & 2.16\,min & 1.13\,min & \cellcolor{bestgreen}\textbf{4.81\,s}
& \cellcolor{failred}1.64 & \cellcolor{failred}52.3 & 2.07 & 8.1 & \cellcolor{bestgreen}\textbf{0.020} \\

\hline
\end{tabular}
}
\end{table*} 
\subsection{Computational Efficiency}
Tables~\ref{tab:biogrid_summary} and~\ref{tab:simfilesize} list the sizes of the input networks and of the similarity files. These topological similarities are precomputed and provided as input to all methods, replacing the sequence or ortholog files that the other methods originally use. Table~\ref{tab:runtime_memory_all} shows the average runtime and peak memory usage over 100 runs of all five methods across species pairs, using Importance and L-GRAAL similarity. DijSLA outperforms all competing methods in both runtime and memory usage for almost every species pair, with either similarity measure. GASOLINE is the second fastest, followed by LePrimAlign. In memory usage, LePrimAlign ranks second with Importance, and GASOLINE ranks second with L-GRAAL. AlignMCL and AlignNemo suffer from out-of-memory (OOM) errors while building the union or alignment graph, and AlignNemo also takes excessive time to run. Their only completed runs are on MM--HS, where the input is small, and these produce no alignments. 

DijSLA’s advantage holds even against the next-best method for each measure, and it widens as the input grows. With Importance similarity, DijSLA typically finishes within 30 seconds per alignment, around $2\times$ faster than GASOLINE, the next-fastest method, and uses only about $1/18$–$1/3$ of the memory of LePrimAlign, the next most memory-efficient. The L-GRAAL similarity file is more than $10\times$ larger, yet DijSLA still finishes within 5 minutes per alignment. It is $2$–$3\times$ faster than GASOLINE for most species pairs and up to $14\times$ faster for RN--HS, while using about $1/100$ of its memory.

\begin{table*}[t]
\centering
\caption{Alignment Size and Quality of the Local Alignment Methods.\protect\\ \upshape Average size ($|\mathcal{A}|$), EC and $S^3$ of the largest alignment over 100 runs for each species pair with Importance or L-GRAAL similarity. EC is measured in the source network, and DijSLA is reported at EC thresholds $\epsilon = 0.1$ and $0.85$. AlignMCL and AlignNemo are omitted, as their only completed runs (MM--HS, Importance) returned empty alignments; n/a marks metrics undefined for empty alignments.}
\label{tab:alignment_quality_all}

\resizebox{\textwidth}{!}{
\begin{tabular}{ll|cccc|cccc|cccc}
\hline

\multirow{3}{*}{\textbf{Species}} &
\multirow{3}{*}{\textbf{Sim.}} &
\multicolumn{4}{c|}{\textbf{$|\mathcal{A}|$}} &
\multicolumn{4}{c|}{\textbf{EC}} &
\multicolumn{4}{c}{\textbf{$S^3$}} \\

&
& \multirow{2}{*}{\makecell{LePrim\\Align}}
& \multirow{2}{*}{GASOLINE}
& \multicolumn{2}{c|}{DijSLA (ours)}
& \multirow{2}{*}{\makecell{LePrim\\Align}}
& \multirow{2}{*}{GASOLINE}
& \multicolumn{2}{c|}{DijSLA (ours)}
& \multirow{2}{*}{\makecell{LePrim\\Align}}
& \multirow{2}{*}{GASOLINE}
& \multicolumn{2}{c}{DijSLA (ours)} \\

&
& & & 0.1 & 0.85
& & & 0.1 & 0.85
& & & 0.1 & 0.85 \\

\hline

\multirow{2}{*}{SC--HS}
& Imp
& 68 & 16 & 1100 & 74
& 0.18 & 0.88 & 0.10 & 0.85
& 0.09 & 0.83 & 0.07 & 0.75 \\

& L-GRAAL
& 17 & 16 & 4768 & 247
& 0.56 & 0.86 & 0.10 & 0.85
& 0.26 & 0.84 & 0.07 & 0.74 \\

\hline

\multirow{2}{*}{DM--HS}
& Imp
& 0 & 7 & 1130 & 105
& n/a & 0.96 & 0.15 & 0.85
& n/a & 0.86 & 0.08 & 0.74 \\

& L-GRAAL
& 7 & 7 & 5620 & 280
& 0.38 & 0.94 & 0.28 & 0.85
& 0.25 & 0.86 & 0.15 & 0.74 \\

\hline

\multirow{2}{*}{CE--HS}
& Imp
& 2 & 5 & 138 & 45
& 1.00 & 0.99 & 0.35 & 0.87
& 1.00 & 0.91 & 0.22 & 0.76 \\

& L-GRAAL
& 24 & 7 & 2185 & 732
& 0.28 & 0.96 & 0.56 & 0.87
& 0.12 & 0.91 & 0.27 & 0.75 \\

\hline

\multirow{2}{*}{AT--HS}
& Imp
& 0 & 9 & 22 & 7
& n/a & 0.94 & 0.39 & 0.96
& n/a & 0.86 & 0.31 & 0.93 \\

& L-GRAAL
& 67 & 10 & 954 & 265
& 0.1 & 0.93 & 0.52 & 0.87
& 0.05 & 0.87 & 0.31 & 0.77 \\

\hline

\multirow{2}{*}{SP--HS}
& Imp
& 6 & 7 & 29 & 16
& 0.88 & 0.96 & 0.71 & 0.91
& 0.47 & 0.93 & 0.54 & 0.84 \\

& L-GRAAL
& 2 & 9 & 844 & 317
& 1.00 & 0.92 & 0.48 & 0.87
& 1.00 & 0.86 & 0.32 & 0.76 \\

\hline

\multirow{2}{*}{MM--HS}
& Imp
& 21 & 0 & 1 & 1
& 0.47 & n/a & n/a & n/a
& 0.12 & n/a & n/a & n/a \\

& L-GRAAL
& 29 & 7 & 82 & 35
& 0.15 & 0.97 & 0.80 & 0.95
& 0.05 & 0.97 & 0.59 & 0.88 \\

\hline

RN--HS
& L-GRAAL
& 10 & 4 & 13 & 6
& 0.73 & 1.00 & 0.81 & 0.99
& 0.64 & 1.00 & 0.67 & 0.97 \\

\hline
\end{tabular}}

\vspace{0.8em}
\includegraphics[width=0.6\textwidth]{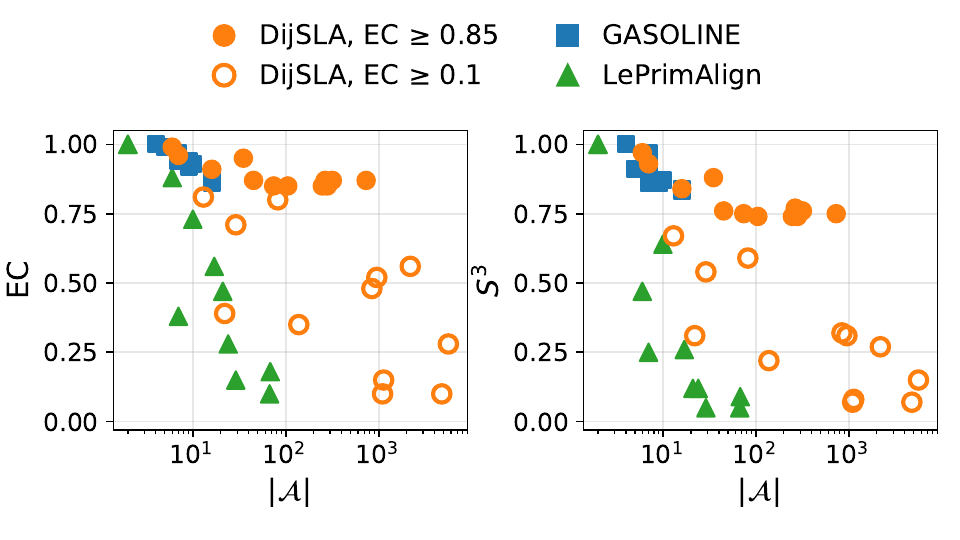}
\captionof{figure}{EC (left) and $S^3$ (right) against alignment size $|\mathcal{A}|$ (log scale) for the alignments in Table~\ref{tab:alignment_quality_all}. Good alignments lie toward the upper right, being both large and topologically similar. DijSLA at $\epsilon=0.85$ is the only method in this region: its EC and $S^3$ are comparable to GASOLINE's, while its alignments are on average larger. At $\epsilon=0.1$, DijSLA trades some EC for alignments of thousands of node pairs. }
\label{fig:size_quality}
\end{table*}

\subsection{Alignment Quality}
In each run, DijSLA produces a single large one-to-one alignment, whereas GASOLINE and LePrimAlign produce many small, separate alignments. We define the \emph{alignment size}, $|\mathcal{A}|$, as the number of aligned node pairs in the largest alignment of a run, and compute EC and $S^3$ on this same alignment. Table~\ref{tab:alignment_quality_all} summarizes the alignment quality of all methods, averaged over 100 runs, in terms of $|\mathcal{A}|$, EC, and $S^3$, and Fig.~\ref{fig:size_quality} shows these results visually. Since high EC is easier to achieve in small alignments, we evaluate DijSLA at two EC thresholds, $\epsilon = 0.1$ and $\epsilon = 0.85$, to enable a fairer comparison of alignment size. 

DijSLA produces far larger alignments than GASOLINE and LePrimAlign on almost every species pair, even at the strict threshold $\epsilon = 0.85$. At $\epsilon = 0.85$, DijSLA's alignments are on average $9.4\times$ larger than GASOLINE's and $4.7\times$ larger than LePrimAlign's (geometric means), and larger in 11 of 12 and 9 of 11 species pairs and similarity combinations, respectively. At $\epsilon = 0.1$, they are $45.5\times$ and $19.0\times$ larger, and larger in all 12 and in 10 of 11 combinations. With L-GRAAL similarity and $\epsilon = 0.1$, we align at least 59.4\% of the proteins of five of the seven species to the human network, including as many as 85.0\% of yeast proteins, with EC ranging from 0.10 to 0.56. For the remaining two species, MM and RN, we align 15.5\% (EC 0.80) and 5.6\% (EC 0.81) of proteins, respectively. Even at $\epsilon = 0.85$, DijSLA aligns 247--732 node pairs on these five species with EC 0.85--0.87, whereas GASOLINE's largest alignments have only 7--16 pairs (EC 0.86--0.96), and LePrimAlign's have 2--67 pairs, mostly with far lower EC (0.10--0.56). AlignMCL and AlignNemo methods are so dependent on sequence-based homology that they cannot use topology alone. We therefore compare function prediction only with LePrimAlign and GASOLINE in the following section.

\begin{figure*}[t]
\centering
\includegraphics[width=\textwidth]{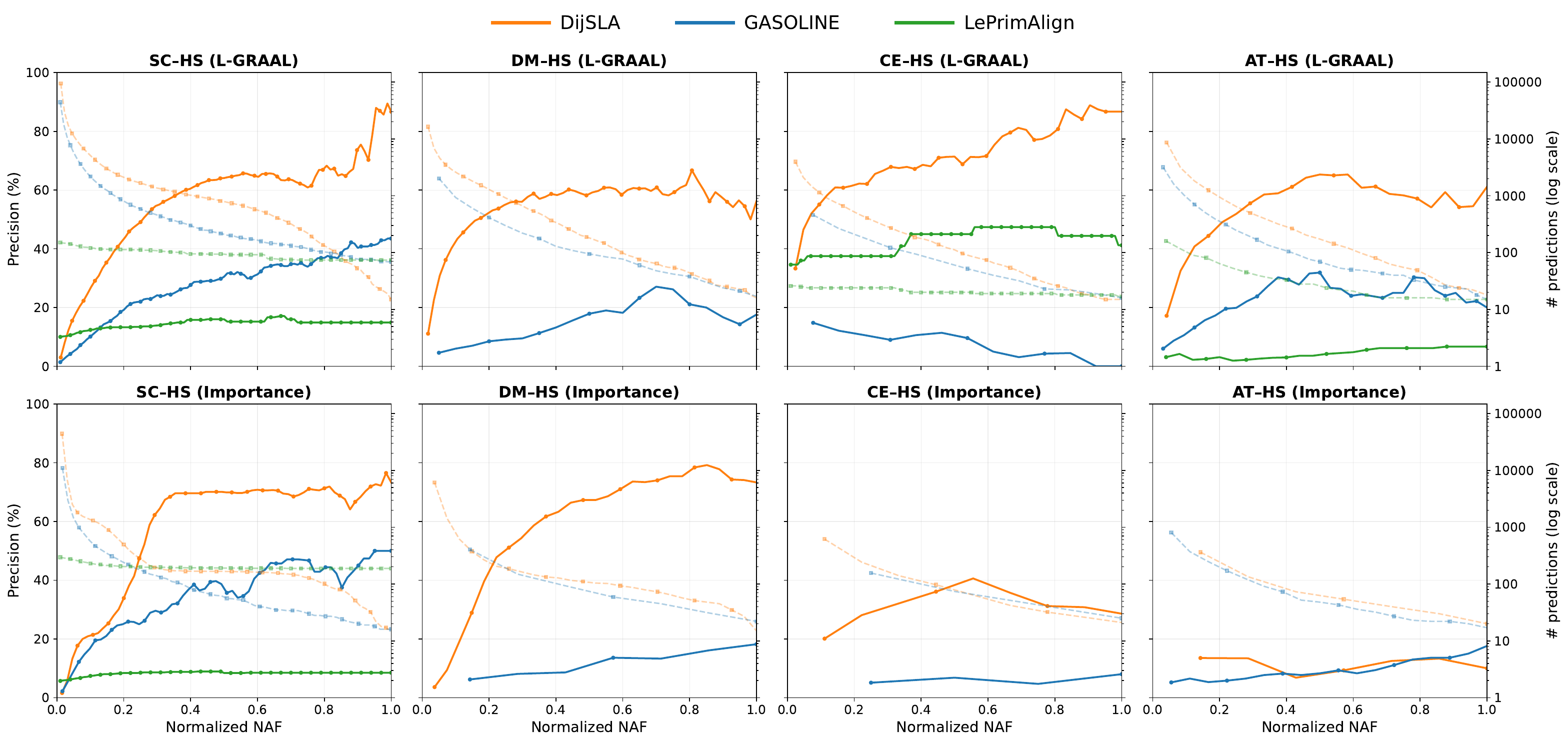}
\caption{Prediction precision (left axis, solid lines) and number of predictions (right axis, dashed lines, log scale) as a function of normalized NAF for DijSLA, GASOLINE, and LePrimAlign, guided by L-GRAAL or Importance. NAF values with fewer than 15 predictions are omitted.}
\label{fig:precision_comparison}
\end{figure*}

\begin{table}[t]
\caption{Sequence-Based GO Evidence Codes Disallowed in Prediction.\protect\\ \upshape We adopt a more restrictive interpretation of``sequence-based'' than standard practice, disallowing every code in which sequence information could have played a role, including manually curated sequence comparisons. }
\label{tab:go_disallowed}
\centering
\begin{tabular}{l p{0.8\columnwidth}}
\toprule
\textbf{Code} & \textbf{Description of sequence-based evidence (disallowed in our predictions)} \\
\midrule
IBA & Curated transfer based on sequence ancestry \\
IEA & Electronic annotation inferred from sequence data \\
ISM & Inferred from sequence model \\
ISA & Inferred from sequence alignment \\
ISO & Inferred from sequence orthology \\
IGC & Inferred from genomic context \\
RCA & Inferred from reviewed computational analysis \\
ISS & Inferred from sequence or structural similarity \\
\bottomrule
\end{tabular}
\end{table}

\begin{table}[t]
\centering
\caption{Correlation Between Prediction Precision and NAF for DijSLA.\protect\\ \upshape Pearson correlation ($\rho$) between IPI prediction precision and NAF threshold $\eta$ at $\epsilon = 0.1$, using only thresholds with at least 15 predictions. $n$: number of thresholds included; --: no alignment available.}
\label{tab:naf_precision_correlation}

\resizebox{\columnwidth}{!}{
\begin{tabular}{l rrrr rrrr}
\toprule
& \multicolumn{4}{c}{\textbf{Imp}} & \multicolumn{4}{c}{\textbf{L-GRAAL}} \\
\cmidrule(lr){2-5} \cmidrule(lr){6-9}
\textbf{Species} & $n$ & $\rho$ & $p$ & $z$ & $n$ & $\rho$ & $p$ & $z$ \\
\midrule
SC--HS & 65 & 0.79 & $4.75\times10^{-15}$ & 7.75 & 88 & 0.88 & $1.13\times10^{-29}$ & 11.25 \\
DM--HS & 27 & 0.86 & $5.99\times10^{-9}$ & 5.70 & 57 & 0.60 & $7.53\times10^{-7}$ & 4.81 \\
CE--HS & 9 & 0.31 & $4.14\times10^{-1}$ & 0.22 & 42 & 0.94 & $1.17\times10^{-20}$ & 9.25 \\
AT--HS & 7 & $-0.15$ & $7.51\times10^{-1}$ & $-0.68$ & 24 & 0.59 & $2.23\times10^{-3}$ & 2.84 \\
SP--HS & 4 & 0.94 & $5.92\times10^{-2}$ & 1.56 & 38 & 0.37 & $2.30\times10^{-2}$ & 1.99 \\
MM--HS & -- & -- & -- & -- & 5 & 0.90 & $3.74\times10^{-2}$ & 1.78 \\
RN--HS & -- & -- & -- & -- & 9 & 0.32 & $4.02\times10^{-1}$ & 0.25 \\
\bottomrule
\end{tabular}}
\end{table}

\begin{figure*}[t]
\centering
\includegraphics[width=\textwidth]{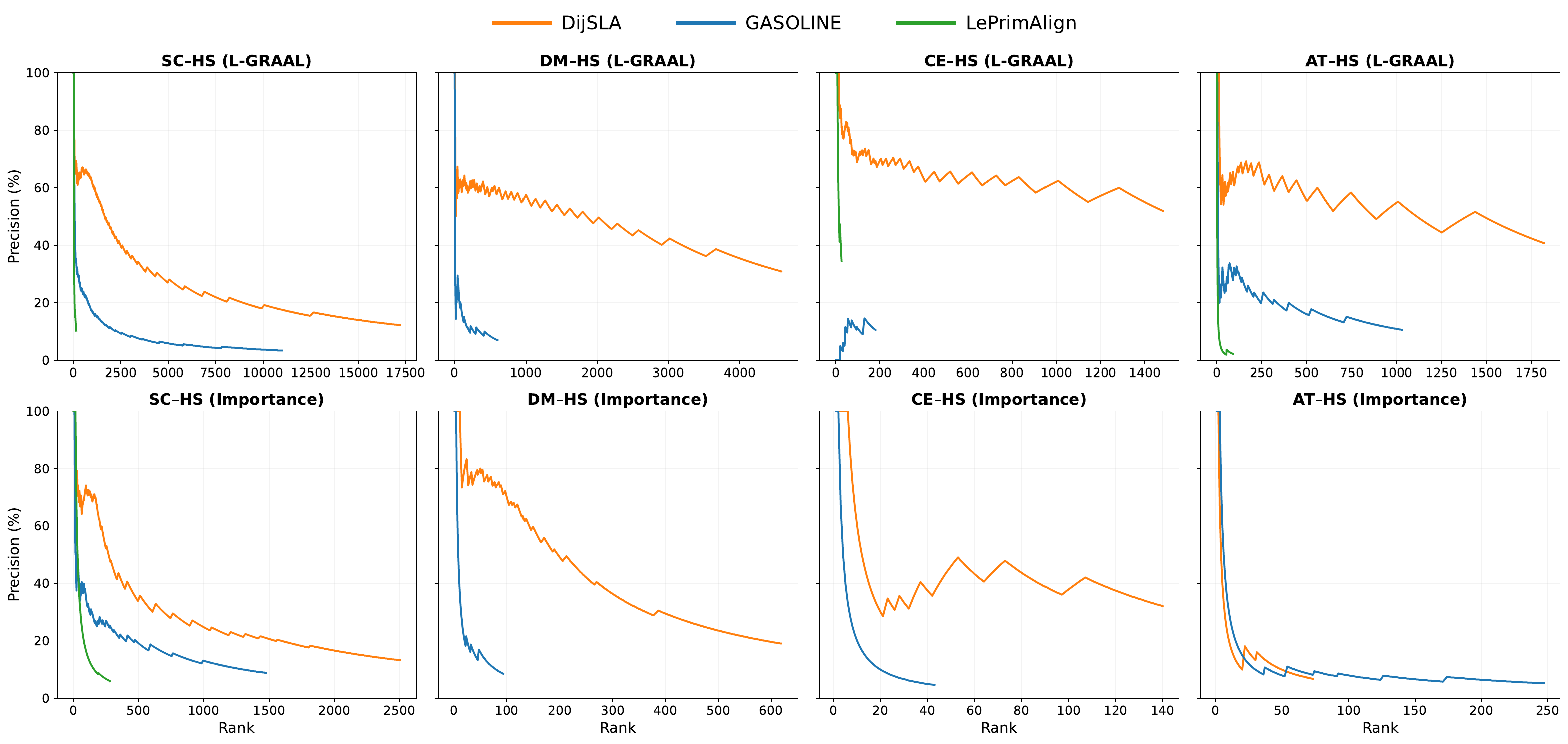}
\caption{To demonstrate the overwhelming number of predictions made by DijSLA, plotted here is the prediction precision as a function of rank, sorted by NAF, for DijSLA (ours), GASOLINE and LePrimAlign. DijSLA usually produces at least an order of magnitude or more predictions at a given prediction precision, compared to GASOLINE and LePrimAlign.}
\label{fig:precision_vs_rank}
\end{figure*}
\subsection{Prediction of Functional Annotations}
\label{sec: GO Pred}

Provided the alignment result, whenever two proteins are aligned and one is annotated with a GO term from April 2010 but the other is not, we transfer the annotation and call it a prediction. We ``wait'' 10 years and verify the prediction against the GO database in April 2020, measuring precision against annotations with the IPI (Inferred from Physical Interaction) evidence code. To predict from topology only, all sequence-based GO evidence codes shown in Table~\ref{tab:go_disallowed} are excluded during prediction, and we exclude any prediction we make that \emph{did} occur with \emph{any} evidence in 2010. This restriction allows us to evaluate whether using network topology alone can identify functionally similar proteins with high precision in the absence of sequence similarity. 

We first evaluate the correlation between prediction precision and NAF across species pairs, using DijSLA with L-GRAAL or Importance as the similarity measure and an EC threshold of $\epsilon = 0.1$. For each NAF threshold $\eta$, we predict GO terms from the aligned pairs with $\mathrm{NAF} \geq \eta$. Thresholds yielding fewer than 15 predictions are excluded as too few for statistically reliable results. Table~\ref{tab:naf_precision_correlation} shows a significant positive correlation between NAF and precision in most species pairs, especially with L-GRAAL, supporting NAF as a confidence signal for the comparisons that follow.

Figure~\ref{fig:precision_comparison} shows prediction precision and the number of predictions as a function of NAF for DijSLA, GASOLINE and LePrimAlign. For a fair comparison across methods, NAF is normalized by the maximum NAF with at least 15 predictions for each method and similarity measure. Precision generally increases with NAF. In all cases except AT--HS with Importance, where all aligners show low precision, DijSLA achieves the highest precision (solid lines), often reaching 80\%--90\% at high NAF and staying above 60\% over a broad range of NAF thresholds. It also makes substantially more predictions (dashed lines). In contrast, the competing methods' precision stays between 20\% and 40\%, with fewer predictions.

LePrimAlign produces no alignments for DM--HS and AT--HS with Importance, and yields fewer than 15 predictions for DM--HS with L-GRAAL and CE--HS with Importance. Its precision also increases less with NAF. Using NAF as a confidence signal requires sufficient randomness in the aligner, so that pairs recurring across runs reflect a real signal rather than repetition. LePrimAlign frequently aligns the same subregion across runs (Table~\ref{tab:alignment_jaccard}), so its NAF largely reflects this deterministic behavior. This may suggest a limitation of methods that start from a global alignment: the initial mapping may restrict them to similar regions in every run, overlooking locally similar regions that conflict with it.
\begin{table}[t]
\centering
\caption{Overlap Between Alignments Across Runs.\protect\\ \upshape Average pairwise Jaccard similarity between alignments across 100 independent runs; higher values indicate that the same node pairs are aligned more often across runs. --: no alignment available.}
\label{tab:alignment_jaccard}
\small
\setlength{\tabcolsep}{3pt}
\begin{tabular}{llcccc}
\toprule
\textbf{Species} & \textbf{Sim.}
& \textbf{GAS} & \textbf{LP}
& \multicolumn{2}{c}{\textbf{DijSLA (ours)}} \\
\cmidrule(lr){5-6}
& & & & $\epsilon=0.1$ & $\epsilon=0.85$ \\
\midrule

\multirow{2}{*}{SC--HS}
& Imp     & 0.025 & 0.846    & 0.001 & 0.003 \\
& L-GRAAL & 0.012 & 0.778    & 0.021 & 0.003 \\
\midrule

\multirow{2}{*}{DM--HS}
& Imp     & 0.025 & --    & 0.004 & 0.002 \\
& L-GRAAL & 0.020 & 0.653 & 0.025 & 0.003 \\
\midrule

\multirow{2}{*}{CE--HS}
& Imp     & 0.018 & 0.147 & 0.011 & 0.005 \\
& L-GRAAL & 0.019 & 0.770 & 0.020 & 0.001 \\
\midrule

\multirow{2}{*}{AT--HS}
& Imp     & 0.021 & --    & 0.018 & 0.014 \\
& L-GRAAL & 0.014 & 0.156    & 0.003 & 0.000 \\
\midrule

\multirow{2}{*}{SP--HS}
& Imp     & 0.003 & 0.727 & 0.007 & 0.008 \\
& L-GRAAL & 0.016 & 0.121 & 0.003 & 0.001 \\
\midrule

\multirow{2}{*}{MM--HS}
& Imp     & --    & 0.172 & 0.203 & 0.203 \\
& L-GRAAL & 0.015 & 0.239 & 0.001 & 0.001 \\
\midrule

\multirow{1}{*}{RN--HS}
& L-GRAAL & 0.023 & 0.192 & 0.030 & 0.031 \\

\bottomrule
\end{tabular}
\end{table}

We note that the number of predictions as a function of NAF is also plotted in Fig.~\ref{fig:precision_comparison}, but the log scale of the right axis may not make it clear the extent to which DijSLA’s \emph{number} of high-precision predictions completely outclasses the other methods. Thus, in Fig.~\ref{fig:precision_vs_rank}, we illustrate prediction quality by plotting prediction precision as a function of prediction rank, where aligned proteins are ranked according to their NAF values. Across all experiments, the prediction precision of both GASOLINE and LePrimAlign declines drastically---indicating that only very few top predictions made by their aligned proteins are validated, while the remaining predictions have low validation rates. In contrast, the precision curve of DijSLA declines much more gradually, demonstrating that DijSLA’s high-precision predictions vastly outnumber those of the other methods. Taking the experiment SC--HS L-GRAAL for example, up to the 100th prediction, the prediction precision of DijSLA remains as high as 67\% while that of GASOLINE slumps to 38\% and that of LePrimAlign plunges to 15\%. Until the 10,000th prediction, DijSLA maintains precision above 19\%, whereas GASOLINE declines below 19\% at its 850th prediction (more than $10\times$ earlier in rank) and LePrimAlign plummets below the same precision at its 59th prediction (almost $170\times$ earlier in rank). Moreover, DijSLA produces significantly more predictions than GASOLINE and LePrimAlign, resulting in a rank span that is at least $1.5\times$ that of GASOLINE and approximately $100\times$ that of LePrimAlign. Notably, despite producing substantially more predictions, the precision of DijSLA at its last rank remains higher than that of the other methods at their last ranks. The precision held by the last rank of DijSLA is equal to that held by only the top 15\% of GASOLINE’s predictions (up to the 1,671st prediction) and the top 85\% of LePrimAlign’s predictions (up to the 124th prediction). In other words, all of DijSLA's predictions match the top 15\% and 85\% of the other methods' predictions in quality, with $10\times$ and $100\times$ more predictions. This observation is not limited to the SC--HS L-GRAAL experiment; similar trends can also be observed in the other experiments.

\begin{figure}[t]
    \centering
    \includegraphics[
        alt={5 different color of lines representing the prediction precision (solid lines) and the number of prediction (dashed lines) with five different delta: 0, 0.01, 0.05, 0.3, 1},
        width=\columnwidth
    ]{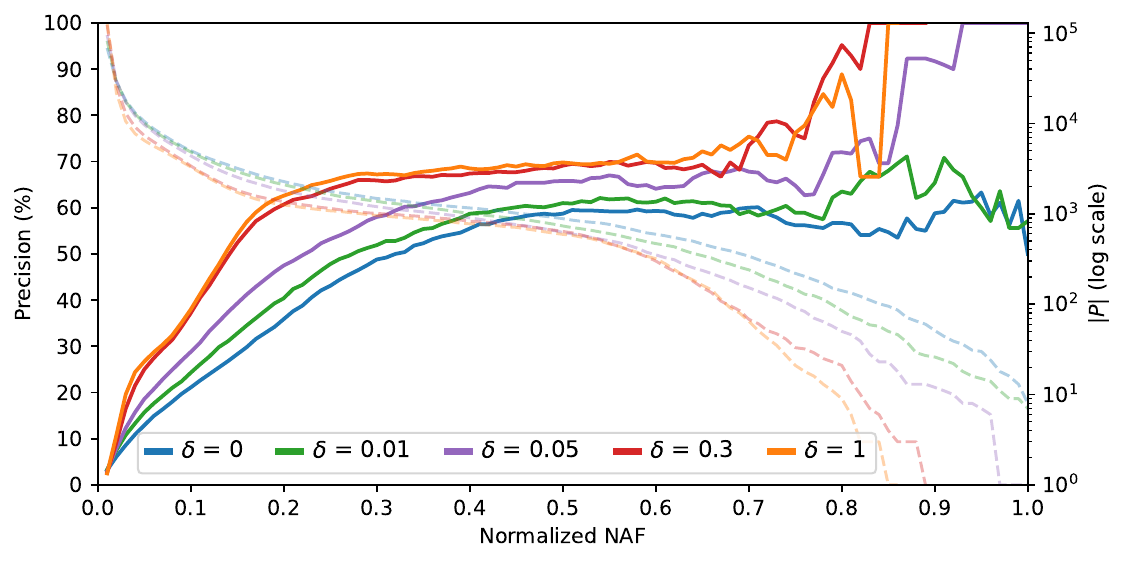}
    \caption{Impact of different $\delta$ values on the number of predictions made from the aligned pairs (dashed lines) and corresponding prediction precision (solid lines) across normalized NAF thresholds for the SC--HS species pair.}
    \label{fig:delta-impact}
\end{figure}

\begin{figure}[t]
    \centering
    \includegraphics[alt={Mean EC value across 100 runs versus the number of aligned pairs, with 5 different color representing five different delta: 0, 0.01, 0.05, 0.3, 1. Mean S3 value across 100 runs versus the number of aligned pairs, with 5 different color representing five different delta: 0, 0.01, 0.05, 0.3, 1}, width=\columnwidth]{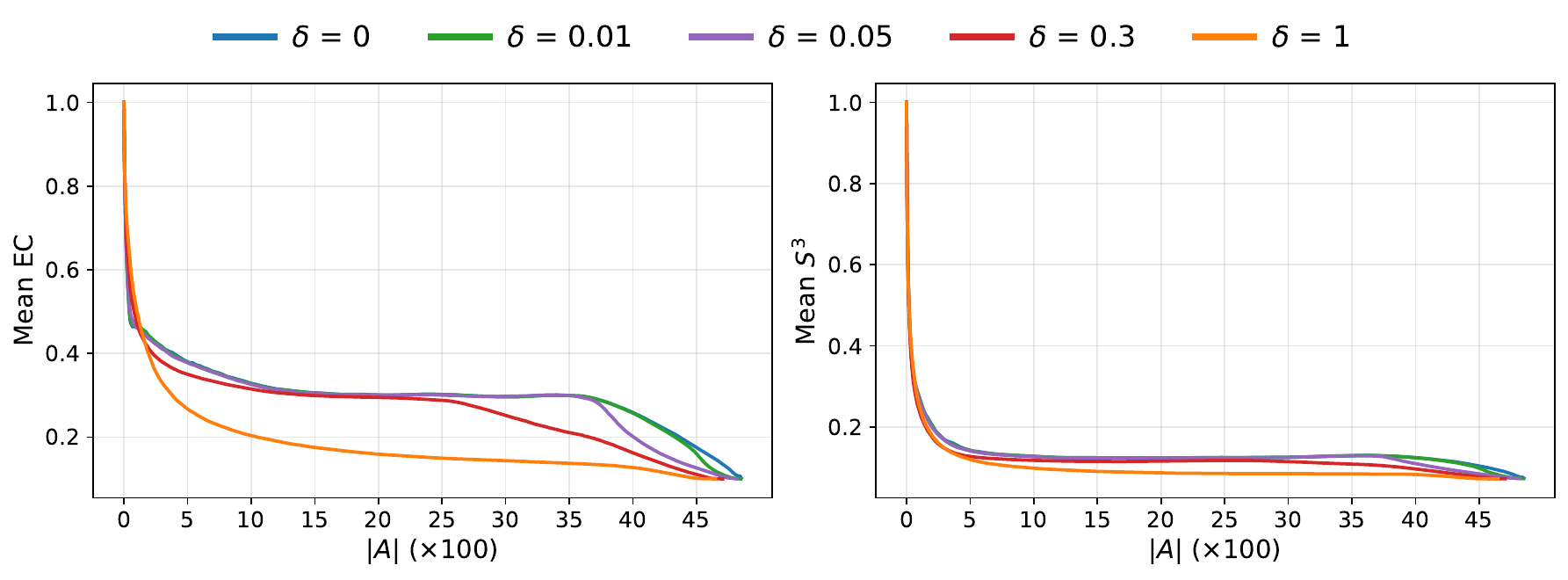}
    \caption{Mean EC (left) and mean $S^3$ (right) as functions of $|\mathcal{A}|$ (×100) for the \textit{S.\ cerevisiae}–\textit{H.\ sapiens} network pair under different values of $\delta$. Each curve shows the mean score averaged over 100 runs at each alignment step, illustrating how increasing $\delta$ influences alignment quality as the number of aligned pairs grows.}
    \label{fig:delta_EC_S3}
\end{figure}

\subsection{Effect of Randomness (\texorpdfstring{$\delta$}{delta})}
To understand how the controlled randomness $\delta$ affects DijSLA, we vary $\delta$ on SC--HS, the most complete network pair in our study, and analyze 1) the number and precision of predictions, and 2) alignment quality, measured by EC and $S^3$. As shown in Fig.~\ref{fig:delta-impact}, larger values of $\delta$ yield higher precision than smaller values at a similar number of predictions, indicating that controlled randomness leads to better alignments. For $\mathrm{NAF} \leq 0.6$, the number of predictions is similar across values of $\delta$, with larger $\delta$ giving slightly fewer predictions but higher precision. The advantage of $\delta \geq 0.3$ is largest for NAF below 0.3 and above 0.6, and smaller but still present in between. This suggests that substantial amounts of controlled randomness lead to substantially \emph{better} alignments, with topological quality maintained by the EC threshold. When the NAF exceeds 0.6, the number of predictions decreases much more sharply, and the effect of $\delta$ on precision becomes more significant. In particular, for $\delta \geq 0.05$, the prediction precision reaches 100\% while the precision for those $\delta \le 0.01$ remain around 60\%. When comparing settings with approximately the same number of predictions, those with $\delta \geq 0.05$ still exhibit higher prediction precision. This indicates that allowing more flexibility in building the alignment (in this case via higher values of $\delta$) may be more important than greedily choosing the local best choice.

As shown in Fig.~\ref{fig:delta_EC_S3}, higher $\delta$ does not substantially reduce EC during the early stages of alignment (approximately the first 3,000 aligned pairs), except for $\delta = 1$, which shows a sharp drop. At later stages, EC decreases more noticeably for larger $\delta$, as expected from the added randomness, but remains above the threshold $\epsilon = 0.1$. $S^3$ changes little across values of $\delta$, likely because it is not explicitly controlled; an $S^3$ threshold similar to the EC threshold could improve it in future work.

\section{Conclusion}

The promise of PPI network topology for protein function prediction has been hard to verify, and very few global aligners have shown promising results from topology alone. In this work, we propose DijSLA, a Dijkstra-inspired stochastic local alignment, and show that topology alone can be used effectively for function prediction. DijSLA produces much larger alignments than competing local network alignment methods, yielding tens of thousands of predictions with high validation rates---orders of magnitude more than existing approaches, while also achieving much higher precision. In addition, DijSLA runs substantially faster and uses remarkably little memory, which makes it scalable to larger biological networks. Compared with the global topological aligner SANA \cite{wang2022sana}, DijSLA achieves broadly similar results: for most species pairs, prediction precision at the highest NAF is well above 50\% over the same 10-year period, with a similar range in the number of predictions.

Currently, seed selection relies on either the top-scoring pair or a random choice among near-top candidates under the similarity measure. Future work could explore alternative seeding strategies and numbers of initial seeds, and test whether they improve alignment robustness or biological interpretability. The other graphlet-based variants terminate after a single alignment step, suggesting high sensitivity to the choice of seed; alternative seeding strategies may therefore yield better alignments, and our companion paper~\cite{wang2026topology} on generating effective seeds may offer significant advantages, though testing this is left for future work. Moreover, our current experiments were conducted using snapshots of the interactome from 2010 and 2020. A valuable extension would be to evaluate performance on more recent, more complete PPI networks and perform forward prediction on the most recent interactome and await future experimental validations as ground truth.

\section*{Acknowledgments}
We thank Nivedha Aravindmali for carefully proofreading the manuscript, and OpenLab at the University of California, Irvine, for providing computing resources.

\bibliographystyle{IEEEtran}
\bibliography{citation}

\end{document}